\documentclass[prl,preprint,showpacs,amssymb,superscriptaddress]{revtex4}
\usepackage{graphicx}% Include figure files
\usepackage{dcolumn}% Align table columns on decimal point
\usepackage{bm}% bold math
\providecommand{\PNNMEAS}{\mbox {$1.47$}} % central value br in units of e-10
\providecommand{\PNNBR}{\mbox {$({\PNNMEAS}^{+1.30}_{-0.89}) \times 10^{-10}$}} % E949&E787 combined BR
\providecommand{\PNNNINETY}{\mbox {$(0.42,3.22)\times10^{-10}$}} % 90%CL interval on E949&E787 combined BR
\providecommand{\PNNNINETYFIVE}{\mbox {$(0.27,3.84)\times10^{-10}$}} % 95%CL interval on E949&E787 combined BR
\providecommand{\PNNOLD}{\mbox {$(1.57^{+1.75}_{-0.82})\times10^{-10}$}} % E787 BR
\providecommand{\PNNSM}{\mbox {$(0.80\pm0.11)\times 10^{-10}$}}  % SM BR
\providecommand{\FEONE}{\mbox {$50$}} % S/B of cell with 1995 candidate
\providecommand{\FETWO}{\mbox {$7$}} % S/B of cell with 1998 candidate
\providecommand{\FCAND}{\mbox {$0.9$}} % S/B  of cell with 2002 candidate
\providecommand{\WTONE}{\mbox {$0.98$}} % Sm/B/(Sm/B+1) for 1995 cand. Sm = 1.38/0.77*S
\providecommand{\WTTWO}{\mbox {$0.88$}} % Sm/B/(Sm/B+1) for 1998 cand. Sm = 1.38/0.77*S
\providecommand{\WTCAND}{\mbox {$0.48$}} % Sm/B/(Sm/B+1) for 2002 cand. Sm = 1.38/0.77*S
\providecommand{\FBONE}{\mbox {$0.006$}} % Integrated background 1995 cand.
\providecommand{\FBTWO}{\mbox {$0.02$}} % Integrated background 1998 cand.
\providecommand{\FBCAND}{\mbox {$0.07$}} % Integrated background 2002 cand.
\providecommand{\BNEW}{\mbox {$0.30$}} % Total background 2002.
\providecommand{\BOLD}{\mbox {$0.14$}} % Total background 2002.
\providecommand{\FAMILONLIMIT}{\mbox {$ 0.73 \times 10^{-10}$}} % 90% CL limit on K->pi,X

\begin{document}

%\date{PostPRL DRAFT1-\today}

\title{Improved Measurement of the  $K^+ \to \pi^+ \nu \bar\nu$ Branching Ratio}

% 16jan04 new authorlist from sk
% 22jan04 mods to authorlist YK=Yuri, PC=Peter

\author{V.V.~Anisimovsky}\affiliation{Institute for Nuclear Research RAS, 60 October Revolution Prospect 7a, 117312 Moscow, Russia}
\author{A.V.~Artamonov}\affiliation{Institute for High Energy Physics, Protvino, Moscow Region, 142 280, Russia}
\author{B.~Bassalleck}\affiliation{Department of Physics and Astronomy, University of New Mexico, Albuquerque, NM 87131}
\author{B.~Bhuyan}\altaffiliation{Also at the Department of Physics, University of Delhi, Delhi 1100007, India} \altaffiliation{Present address: Department of Physics and Astronomy, University of Victoria, Victoria, British Columbia, Canada V8W 3P6.} \affiliation{Brookhaven National Laboratory, Upton, NY 11973}
\author{E.W.~Blackmore}\affiliation{TRIUMF, 4004 Wesbrook Mall, Vancouver, British Columbia, Canada V6T 2A3}
\author{D.A.~Bryman} \affiliation{Department of Physics and Astronomy, University of British Columbia, Vancouver, British Columbia, Canada V6T 1Z1}
\author{S.~Chen} \affiliation{TRIUMF, 4004 Wesbrook Mall, Vancouver, British Columbia, Canada V6T 2A3}
\author{I-H.~Chiang} \affiliation{Brookhaven National Laboratory, Upton, NY 11973}
\author{I.-A.~Christidi} \affiliation{Department of Physics and Astronomy, Stony Brook University, Stony Brook, NY 11794}
\author{P.S.~Cooper}\affiliation{Fermi National Accelerator Laboratory, Batavia, IL 60510}
\author{M.V.~Diwan} \affiliation{Brookhaven National Laboratory, Upton, NY 11973}
\author{J.S.~Frank} \affiliation{Brookhaven National Laboratory, Upton, NY 11973}
\author{T.~Fujiwara}\affiliation{Department of Physics, Kyoto University, Sakyo-ku, Kyoto 606-8502, Japan}
\author{J.~Hu} \affiliation{TRIUMF, 4004 Wesbrook Mall, Vancouver, British Columbia, Canada V6T 2A3}
\author{A.P.~Ivashkin}\affiliation{Institute for Nuclear Research RAS, 60 October Revolution Prospect 7a, 117312 Moscow, Russia}
\author{D.E.~Jaffe} \affiliation{Brookhaven National Laboratory, Upton, NY 11973}
\author{S.~Kabe} \affiliation{High Energy Accelerator Research Organization (KEK), Oho, Tsukuba, Ibaraki 305-0801, Japan}
\author{S.H.~Kettell} \affiliation{Brookhaven National Laboratory, Upton, NY 11973}
\author{M.M.~Khabibullin}\affiliation{Institute for Nuclear Research RAS, 60 October Revolution Prospect 7a, 117312 Moscow, Russia}
\author{A.N.~Khotjantsev}\affiliation{Institute for Nuclear Research RAS, 60 October Revolution Prospect 7a, 117312 Moscow, Russia}
\author{P.~Kitching} \affiliation{Centre for Subatomic Research, University of Alberta, Edmonton, Canada T6G 2N5}
\author{M.~Kobayashi} \affiliation{High Energy Accelerator Research Organization (KEK), Oho, Tsukuba, Ibaraki 305-0801, Japan}
\author{T.K.~Komatsubara} \affiliation{High Energy Accelerator Research Organization (KEK), Oho, Tsukuba, Ibaraki 305-0801, Japan}
\author{A.~Konaka} \affiliation{TRIUMF, 4004 Wesbrook Mall, Vancouver, British Columbia, Canada V6T 2A3}
\author{A.P.~Kozhevnikov}\affiliation{Institute for High Energy Physics, Protvino, Moscow Region, 142 280, Russia}
\author{Yu.G.~Kudenko}\affiliation{Institute for Nuclear Research RAS, 60 October Revolution Prospect 7a, 117312 Moscow, Russia} % YK
\author{A.~Kushnirenko} \altaffiliation{Present address: Institute for High Energy Physics, Protvino, Moscow Region, 142 280, Russia.}  \affiliation{Fermi National Accelerator Laboratory, Batavia, IL 60510} %PC
\author{L.G.~Landsberg}\affiliation{Institute for High Energy Physics, Protvino, Moscow Region, 142 280, Russia}
\author{B.~Lewis}\affiliation{Department of Physics and Astronomy, University of New Mexico, Albuquerque, NM 87131}
\author{K.K.~Li}\affiliation{Brookhaven National Laboratory, Upton, NY 11973}
\author{L.S.~Littenberg} \affiliation{Brookhaven National Laboratory, Upton, NY 11973}
\author{J.A.~Macdonald} \altaffiliation{Deceased.} \affiliation{TRIUMF, 4004 Wesbrook Mall, Vancouver, British Columbia, Canada V6T 2A3}
\author{J.~Mildenberger} \affiliation{TRIUMF, 4004 Wesbrook Mall, Vancouver, British Columbia, Canada V6T 2A3}
\author{O.V.~Mineev}\affiliation{Institute for Nuclear Research RAS, 60 October Revolution Prospect 7a, 117312 Moscow, Russia}
\author{M. Miyajima} \affiliation{Department of Applied Physics, Fukui University, 3-9-1 Bunkyo, Fukui, Fukui 910-8507, Japan}
\author{K.~Mizouchi}\affiliation{Department of Physics, Kyoto University, Sakyo-ku, Kyoto 606-8502, Japan}
\author{V.A.~Mukhin}\affiliation{Institute for High Energy Physics, Protvino, Moscow Region, 142 280, Russia}
\author{N.~Muramatsu} \altaffiliation{Present address: Department of Physics, Kyoto University, Sakyo-ku, Kyoto 606-8502, Japan.}
\affiliation{Research Center for Nuclear Physics, Osaka University, 10-1 Mihogaoka, Ibaraki, Osaka 567-0047, Japan}
\author{T.~Nakano}\affiliation{Research Center for Nuclear Physics, Osaka University, 10-1 Mihogaoka, Ibaraki, Osaka 567-0047, Japan}
\author{M.~Nomachi}\affiliation{Laboratory of Nuclear Studies, Osaka University, 1-1 Machikaneyama, Toyonaka, Osaka 560-0043, Japan}
\author{T.~Nomura}\affiliation{Department of Physics, Kyoto University, Sakyo-ku, Kyoto 606-8502, Japan}
\author{T.~Numao} \affiliation{TRIUMF, 4004 Wesbrook Mall, Vancouver, British Columbia, Canada V6T 2A3}
\author{V.F.~Obraztsov}\affiliation{Institute for High Energy Physics, Protvino, Moscow Region, 142 280, Russia}

\author{K.~Omata}\affiliation{High Energy Accelerator Research Organization (KEK), Oho, Tsukuba, Ibaraki 305-0801, Japan}
\author{D.I.~Patalakha}\affiliation{Institute for High Energy Physics, Protvino, Moscow Region, 142 280, Russia}
\author{S.V.~Petrenko}\affiliation{Institute for High Energy Physics, Protvino, Moscow Region, 142 280, Russia}
\author{R.~Poutissou} \affiliation{TRIUMF, 4004 Wesbrook Mall, Vancouver, British Columbia, Canada V6T 2A3}
\author{E.J.~Ramberg}\affiliation{Fermi National Accelerator Laboratory, Batavia, IL 60510}%PC
\author{G.~Redlinger} \affiliation{Brookhaven National Laboratory, Upton, NY 11973}
\author{T.~Sato} \affiliation{High Energy Accelerator Research Organization (KEK), Oho, Tsukuba, Ibaraki 305-0801, Japan}
\author{T.~Sekiguchi}\affiliation{High Energy Accelerator Research Organization (KEK), Oho, Tsukuba, Ibaraki 305-0801, Japan}
\author{T.~Shinkawa} \affiliation{Department of Applied Physics, National Defense Academy, Yokosuka, Kanagawa 239-8686, Japan}
\author{R.C.~Strand} \affiliation{Brookhaven National Laboratory, Upton, NY 11973}
\author{S.~Sugimoto} \affiliation{High Energy Accelerator Research Organization (KEK), Oho, Tsukuba, Ibaraki 305-0801, Japan}
\author{Y.~Tamagawa} \affiliation{Department of Applied Physics, Fukui University, 3-9-1 Bunkyo, Fukui, Fukui 910-8507, Japan}
\author{R.~Tschirhart}\affiliation{Fermi National Accelerator Laboratory, Batavia, IL 60510}%PC
\author{T.~Tsunemi}\affiliation{High Energy Accelerator Research Organization (KEK), Oho, Tsukuba, Ibaraki 305-0801, Japan}
\author{D.V.~Vavilov}\affiliation{Institute for High Energy Physics, Protvino, Moscow Region, 142 280, Russia}
\author{B.~Viren}\affiliation{Brookhaven National Laboratory, Upton, NY 11973}
\author{N.V.~Yershov}\affiliation{Institute for Nuclear Research RAS, 60 October Revolution Prospect 7a, 117312 Moscow, Russia}
\author{Y.~Yoshimura} \affiliation{High Energy Accelerator Research Organization (KEK), Oho, Tsukuba, Ibaraki 305-0801, Japan}
\author{T.~Yoshioka}\affiliation{High Energy Accelerator Research Organization (KEK), Oho, Tsukuba, Ibaraki 305-0801, Japan}
\collaboration{E949 Collaboration}

\preprint{BNL/72164-2004-JA}
\preprint{KEK/2004-3}
\preprint{TRIUMF/TRI-PP-04-07}

\newpage
\begin{abstract}
An additional event near the upper
kinematic limit for $K^+ \to \pi^+ \nu \bar\nu$
has been observed by Experiment E949 at
Brookhaven National Laboratory. 
%The probability that the background alone gave rise
%to this or any more signal-like configuration was 0.08.
Combining previously reported and new data, 
the branching ratio is ${\cal B}$($K^+ \to \pi^+ \nu
\bar\nu$)$=\PNNBR $ based on three
 events observed in the pion momentum region $211<P<229$ MeV/$c$. 
At the measured central value of the branching ratio, 
the additional event had a signal-to-background ratio of \FCAND. 
\end{abstract}
\pacs{ 13.20.Eb, 12.15.Hh, 14.80.Mz}

\maketitle

In the standard model (SM), the decay $K^+ \to \pi^+ \nu \bar\nu$~ is
 sensitive to the couplings of top quarks which dominate the internal
 processes involved in this flavor changing neutral current reaction. A
 reliable SM prediction for the branching ratio ${\cal B}(K^+ \to
 \pi^+ \nu \bar\nu$)=\PNNSM\ \cite{BB,newbb} can be made due to
 knowledge of the hadronic transition matrix element from
 similar processes, and minimal complications from hadronic
 effects.
%\cite{B2004}.  
${\cal B}(K^+ \to \pi^+ \nu \bar\nu)$ is a
 sensitive probe of new physics, since, for example, the apparent
 couplings between top and down quarks may also be determined by
 measurements in the B meson system resulting in a possible
 discrepancy\cite{B2004,BSM}.
%Fleischer:2003xx,Buras:2002ej,Grossman:1997sk,he04,
% grossman04,buras04,burdmsn02,barenboim01}.
 In earlier studies, two
 events consistent with the decay $K^+ \to \pi^+ \nu \bar\nu$ were
 reported giving  ${\cal B}$($K^+ \to \pi^+ \nu
 \bar\nu$)= \PNNOLD\ by Experiment E787 at the Alternating Gradient
 Synchrotron (AGS) of Brookhaven National
 Laboratory~\cite{pnn2002}.  In this letter, the first
 results from Experiment E949~\cite{det.E949} at the AGS are
 presented.

 Measurement of $K^+ \!  \rightarrow \! \pi^+ \nu \overline{\nu}$ decay from kaons at rest
 involved observation of the $\pi^+$ in the momentum region $211< P
 <229$ MeV/$c$ in the absence of other coincident activity.  Pions were
 identified by comparing momentum ($P$), range ($R$), and energy ($E$)
 measurements, and by observation of the 
$\pi^+ \!  \rightarrow \! \mu^+ \!  \rightarrow \!  e^+$ 
decay sequence.
% at rest. 
%using 500 MHz transientdigitizers\cite{det.TD}. 
Primary background sources were pions from the two-body decay
$K^+\to\pi^+\pi^0$ ($K_{\pi 2}$), muons from $K^+\to\mu^+\nu$ ($K_{\mu 2}$) 
and other $K^+$ decays,
pions scattered from
 the beam, and $K^+$ charge exchange  reactions followed by 
$K_L^0\to\pi^+ l^- \overline{\nu}_l$, where $l=e$ or $\mu$.

The new data were acquired in 2002 using beams, apparatus, and
 procedures  similar to those of  Experiment
 E787~\cite{pnn2002,det.TD,det,det.collar,newl0}. The number of kaons
 stopped in the scintillating fiber target was $N_K=1.8 \times 10^{12}$.
  Measurements of charged decay
 products were made in a 1T magnetic field using the target, a central drift chamber, and a
 cylindrical range stack (RS) of scintillator detectors.  Photons were
 detected in a $4\pi$ sr calorimeter consisting of a lead/scintillator
 sandwich barrel veto  detector (upgraded for E949) 
 surrounding the RS, endcaps of
 undoped CsI crystals, and other detectors. 
   The upgraded apparatus also included replacement of one
 third of the RS, and an improved trigger system\cite{newl0}.
 % additional ancillary photon veto
%  systems~\cite{det.collar}, and an LED flasher system to aid in the RS energy
%  calibration. 
%One third of the RS scintillation counters were
%  replaced to  increase the  light output.
%  An improved trigger system using additional position
% information in the RS allowed  more efficient 
% running and reduced deadtimes at high instantaneous
%   rates\cite{newl0}.  
%As in E787, the analysis criteria  were designed to  yield stable,
%predictable background %rates
%levels of less than one event while achieving 
%maximal sensitivity to ${\cal B}(K^+ \to \pi^+ \nu \bar\nu)$.
Although the instantaneous detector rates were  twice those in E787,
%~\cite{BadAGS}
  detector upgrades and
the use of improved pattern recognition software enabled
comparable acceptance to be obtained.

	Each background source was suppressed by two groups of
complementary but independent selection criteria (cuts), and the
desired level of background rejection was obtained by adjusting the
severity of the cuts. For example, the cut pair for $K_{\pi 2}$
background involved kinematic measurements of the $\pi^+$, and photon
detection in $\pi^0\to\gamma\gamma$ decay.  The photon detection
criteria, for instance, could be varied by changing the energy
threshold and timing coincidence interval (relative to the $\pi^+$
signal) of the photon detectors.  The effectiveness of each cut at
rejecting background was determined using data selected by inverting
the criteria of the complementary cut.  Unbiased estimates of the
effectiveness of the cuts were obtained using a uniformly-sampled 1/3
portion of the data for cut development and the remaining 2/3 portion
for background measurement. Examination of the pre-determined signal
region was avoided throughout the procedure.  The level of signal
acceptance as a function of cut severity was determined using data and
simulations.  This procedure enabled estimates of the expected
background and signal rates inside and outside the signal region at
different levels of background rejection and signal acceptance.

As a check of the method, the observed background levels near but outside the
signal region were compared to the predicted background rates when 
 both cuts for each background type were applied. 
The results  are summarized in Table~\ref{tab:otb} for the
two-body backgrounds, $K_{\pi 2}$ and $K_{\mu 2}$, and the multi-body
background ($K_{\mu m}$) with contributions from
$K^+\to\mu^+\nu\gamma$, $K^+\to\mu^+\pi^0\nu$, and $K_{\pi 2}$ with
$\pi^+\to\mu^+\nu$ decay-in-flight. 
Five cases were  considered corresponding
to increasing  background  levels 
outside the signal region. For example, for the
$K_{\mu 2}$ component, the region nearest to (farthest from) the signal
region was chosen to have 7 (400) times the  background expected in 
the signal region.
The five ratios of
the observed to predicted backgrounds were fitted
to a constant $c$ for each background type. The consistency of $c$
with unity and the acceptable probability of $\chi^2$ of each fit
confirmed both the independence of the pairs of cuts and the
reliability of the background estimates. The measured uncertainties in
the constants c were used to estimate the systematic uncertainties in
the predicted background rates in the signal region.
 
\begin{table}
\begin{tabular}{|l|cll|c|c|}
\hline
Background	&\multicolumn{3}{c|}{$c$}					& $\chi^2$ Probability		& Events		\\
\hline
$K_{\pi 2}$	&$0.85$&${}^{+0.12}_{-0.11}$&${}^{+0.15}_{-0.11}$ 	& 0.17		&$0.216\pm0.023$	\\
$K_{\mu 2}$	&$1.15$&${}^{+0.25}_{-0.21}$&${}^{+0.16}_{-0.12}$	& 0.67		&$0.044\pm0.005$	\\
$K_{\mu m}$    &$1.06$&$ {}^{+0.35 }_{-0.29 }$& ${}^{+0.93}_{-0.34}$		& 0.40		&$0.024\pm0.010$	\\
\hline
\end{tabular}
\caption{\label{tab:otb} The fitted constants $c$ of the ratios of the
observed to the predicted numbers of background events and the
probability of $\chi^2$ of the fits for the $K_{\pi 2}$, $K_{\mu 2}$
and $K_{\mu m}$ backgrounds near but outside the signal region.  The
first uncertainty in $c$ was due to the statistics of the observed
events and the second was due to the uncertainty in the predicted
rate.  The predicted numbers of background events within the signal
region and their statistical uncertainties are also tabulated in the
fourth column.  Other backgrounds contributed an additional
$0.014\pm0.003$ events resulting in a total number of background
events expected in the signal region of $\BNEW \pm 0.03$.}
\end{table}

To estimate  ${\cal B}(K^+ \to \pi^+ \nu\bar\nu)$, the parameter
space of observables in the signal region was subdivided into 3781
bins corresponding to different ranges of cut severity
 and each observed event could be  assigned to the  bin corresponding to its measured quantities.  
Bin $i$ was characterized by the value of $S_i/b_i$, the
relative probability of an event in the bin to originate from $K^+ \to
\pi^+ \nu\bar\nu$ ($S_i$) or background ($b_i$)~\cite{pnn2002}.  The
signal rate of a bin was $S_i\equiv{\cal B}(K^+ \to \pi^+
\nu\bar\nu)A_iN_K$ where $A_i$ was the acceptance of the $i^{\rm th}$
bin. Each observed event could also be described by a weight $W\equiv
S/(S+b)$ that represented its effective contribution to ${\cal B}(K^+
\to \pi^+ \nu\bar\nu)$.  ${\cal B}(K^+ \to \pi^+ \nu\bar\nu)$ was
obtained by a likelihood ratio technique~\cite{Junk} that determined
the confidence level (C.L.) of a given branching ratio based on the
observed events.  For the 2002 data set, the candidate selection
requirements were similar to those used previously. The pre-determined
signal region was enlarged, resulting in 10\% more acceptance but also
allowing more background.   Estimated background levels
dominated by $K_{\pi 2}$ and $K_{\mu 2}$ are listed in
Table~\ref{tab:otb}.

Examination of the signal region for the new data set yielded one
event with $P=227.3 \pm 2.7$~MeV/$c$, $R=39.2 \pm 1.2$~cm (in
equivalent cm of scintillator), and $E=128.9 \pm 3.6$~MeV. The event
(2002A) has all the characteristics of a signal event although 
 its high momentum and low apparent time of $\pi\to\mu$ decay
(6.2 ns) indicate a higher probability  than the two previously
observed candidate events that it was due to background, particularly
$K_{\mu 2}$ decay.
%~\cite{pme}.

\begin{figure}
\includegraphics*[width=\linewidth]{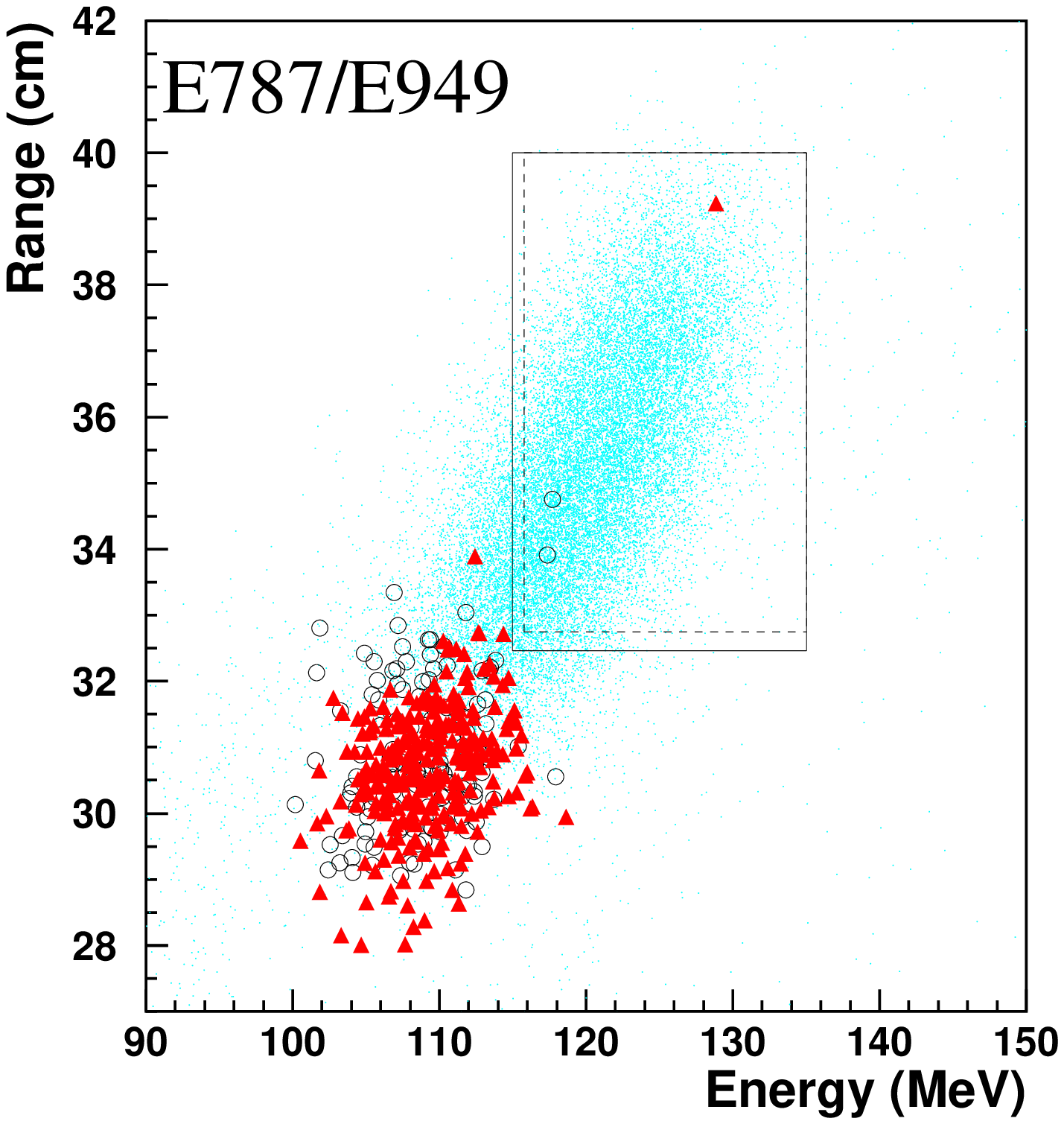}
%{box_new_fullmc.epsi}
\caption{ Range (R)  vs. energy (E) distribution 
of events passing all other cuts of the
final sample. The circles  represent  E787  data  
and the triangles   E949 data.
The group of events around $E=108$~MeV was
due to the $K_{\pi 2}$ background.  The simulated distribution of
%expected 
events from $K^+ \!  \rightarrow \!  \pi^+ \nu\overline{\nu}$~ decay  is indicated by dots.
The solid-line (dashed-line)  box represents the signal region
for E949 (E787).}
\label{rve}
\end{figure}

The combined result for the E949
and E787 data is shown in Figure~\ref{rve} with the range and kinetic
energy of the  events surviving all other cuts.  The result
obtained from the likelihood method described above was ${\cal B}(K^+
\to \pi^+ \nu \overline{\nu}) =\PNNBR\ $ incorporating  the three observed
events and their associated weights $W$ given in
Table~\ref{Nbg}.  
For event 2002A the weight was  $W= \WTCAND$ ($S/b=\FCAND$). 
The estimated probability that the background alone gave rise
to this or any more signal-like event was \FBCAND.
Table~\ref{Nbg} also shows the 
estimated probability that the background alone gave rise
to each event (or any more signal-like event),
the acceptances~\cite{pnn2002},
$N_K$, and the total expected background levels.  
This result is consistent with the SM expectation~\cite{BB,newbb}.  The
quoted 68\% C.L. interval includes statistical and estimated systematic
uncertainties.  The 80\% and 90\% C.L.  intervals for ${\cal
B}(K^+\to\pi^+\nu\bar\nu)$ were \PNNNINETY\ and \PNNNINETYFIVE,
respectively\cite{cl}.
 The estimated systematic uncertainties do not
significantly affect the  confidence levels.
 The estimated probability that
background alone gave rise to the
three observed events (or to  any more signal-like configuration) 
was $0.001$\cite{E787redux0}.

\begin{table}
\begin{tabular}{|l|c|c|c|}
\hline
~			&\multicolumn{2}{c|}{E787} 			&E949\\
\hline
$N_{K} $ 	 	&\multicolumn{2}{c|}{$5.9 \times10^{12}$}& $1.8\times 10^{12}$\\
\hline
Total Acceptance 	&\multicolumn{2}{c|}{$0.0020 \pm 0.0002 $}	& $ 0.0022\pm 0.0002 $ \\ 
\hline 
Total Background 	&\multicolumn{2}{c|}{$\BOLD \pm 0.05$} 		& $\BNEW \pm 0.03$\\
\hline
Candidate		& 1995A		& 1998C				& 2002A \\
%\hline
%Number of Candidates 	& 1	& 1	&1  \\
\hline
$S/b$  & \FEONE\ & \FETWO\ & \FCAND\ \\
\hline
$ W$     & \WTONE\ & \WTTWO\ & \WTCAND\ \\
\hline 
Background Prob.   	& \FBONE\ 	& \FBTWO\ 			& \FBCAND\ \\
\hline 
\end{tabular}
\caption{\label{Nbg} Numbers of kaons stopped in the target $N_{K}$,
total acceptance, total numbers of estimated background events for the
E949 and E787 data samples\cite{E787redux}, $S/b$ and $W$ for
%the bin containing % added 28apr04
each observed candidate event calculated from the likelihood analysis
described in the text, and the 
estimated probability that the background alone gave rise
to each event (or any more signal-like event).
}
\end{table}

The  E787 and E949 data  were also 
used to set a  limit on the branching ratio  
for  $K^+ \!\rightarrow \!  \pi^+ X^0$, where $X^0$ is a neutral weakly
interacting massless particle~\cite{x0}.  Previous E787 data produced
a limit of 
${\cal B}(K^+ \!  \rightarrow \!\pi^+ X^0) < 0.59 \times 10^{-10}$\cite{pnn2002}.
 The new result 
was ${\cal B}(K^+ \!  \rightarrow \!\pi^+ X^0) < \FAMILONLIMIT\ $ (90\% C.L.), based on 
the inclusion of event 2002A %the single event described above 
which was 
observed within two standard deviations of the expected pion momentum.

We 
acknowledge the dedicated effort of the technical staff supporting
E949, the Brookhaven C-A Department,
and  the  contributions made by colleagues who
participated in  E787.  This research
was supported in part by the U.S. Department of Energy, 
%under Contracts No. DE-AC02-98CH10886, 
the Ministry of Education, Culture, Sports, Science and Technology of
Japan through the Japan-U.S. Cooperative Research Program in High
Energy Physics and under Grant-in-Aids for Scientific Research, the
Natural Sciences and Engineering Research Council and the National
Research Council of Canada, the Russian Federation 
State Scientific Center Institute for High
Energy Physics, and the Ministry of Industry, Science and
New Technologies of the Russian Federation.


\begin{thebibliography}{ (a) }
\bibitem{E787redux} A corrected  estimate of the  
number of background events expected in the E787 data sample due to a reevaluation of the beam background
is presented here.
\end{thebibliography}
\end{document}